# Comment on "Dual-wavelength in-line phase-shifting interferometry based on two dc-term-suppressed intensities with a special phase shift for quantitative phase extraction"

Minmin Wang[a], Guangliang Du[a], Canlin Zhou*[a], Chaorui Zhang[a], Shuchun Si[a], Hui Li[a], Zhenkun Lei[b], YanJie Li[c]

(a. School of Physics, Shandong University, Jinan , 250100, China
b. Department of engineering mechanics, Dalian University of Technology, Dalian,116024
c. School of civil engineering and architecture, Jinan University, Jinan, 2250020 )
*Corresponding author: Tel: +8613256153609;     E-mail address: canlinzhou@sdu.edu.cn;
**Corresponding author: Tel: +8615841175236;    E-mail address: leizk@163.com

We comment on the recent Letter by Xu and Wang et al. [Opt. Lett. 41, 2430 (2016)] in which an approach of quantitative phase extraction in dual-wavelength in-line phase-shifting interferometry (DWILPSI) was proposed. It is noted that a special phase shift is used, which more or less embarrasses its practical operation. We wish to show that the same result can also be reached by combining the generalized phase-shifting algorithm and the least-square algorithm, in which the phase shift can be chosen randomly. In addition to maintaining high accuracy and rapid processing speed of the DWILPSI method, the proposed method greatly facilitates its application in actual measurement.

In a recent Letter Xu and Wang et al.[1] presented an approach of quantitative phase extraction in dual-wavelength in-line phase-shifting interferometry (DWILPSI) based on two intensities without the corresponding dc terms for each wavelength, in which a special phase shift is employed. With the phase retrieved, one can readily reconstruct the thickness of the phase object. The main purpose of this Letter is not to discuss the reasonability or to point out some errors. In fact, we want to improve the method based on the generalized phase-shifting algrithom[2] and the least-square algorithm[3,4], so that the special phase shift is no longer needed, bringing much convenience in actual operation.

An analysis of the the proposed method follows. Eq. (1) of Ref. 1 can be rewritten as

$$I_n(x,y) = a(x,y) + b_{\lambda1}(x,y)\cos[\varphi_{\lambda1}(x,y)]\cos(\delta_{\lambda1,n})$$
$$- b_{\lambda1}(x,y)\sin[\varphi_{\lambda1}(x,y)]\sin(\delta_{\lambda1,n}) + b_{\lambda2}(x,y)\cos[\varphi_{\lambda2}(x,y)]\cos(\delta_{\lambda2,n}) \quad (1)$$
$$- b_{\lambda2}(x,y)\sin[\varphi_{\lambda2}(x,y)]\sin(\delta_{\lambda2,n})$$

Instead of making the subtraction operation between the intensity distribution of two interferograms[1], we define a new set of variables as $b(x,y) = b_{\lambda1}(x,y)\cos[\varphi_{\lambda1}(x,y)]$, $c(x,y) = -b_{\lambda1}(x,y)\sin[\varphi_{\lambda1}(x,y)]$, $d(x,y) = b_{\lambda2}(x,y)\cos[\varphi_{\lambda2}(x,y)]$, and $e(x,y) = -b_{\lambda2}(x,y)\sin[\varphi_{\lambda2}(x,y)]$, Eq. (1) can be described as

$$I_n(x,y) = a(x,y) + b(x,y)\cos(\delta_{\lambda 1,n}) + c(x,y)\sin(\delta_{\lambda 1,n}) \\ + d(x,y)\cos(\delta_{\lambda 2,n}) + e(x,y)\sin(\delta_{\lambda 2,n}) \tag{2}$$

The deviation square sum for all $N$ fringe patterns can be expressed as

$$E(x,y) = \sum_{n=1}^{N} \{a(x,y) + b(x,y)\cos(\delta_{\lambda 1,n}) + c(x,y)\sin(\delta_{\lambda 1,n}) \\ + d(x,y)\cos(\delta_{\lambda 2,n}) + e(x,y)\sin(\delta_{\lambda 2,n}) - I_n'(x,y)\}^2 \tag{3}$$

Where $I_n'$ is the intensity of interferograms measured in the experiment. The phase shifts of two reference waves are produced simultaneously by a PZT, which are already known, while $a(x,y)$, $b(x,y)$, $c(x,y)$, $d(x,y)$, $e(x,y)$ are left as unknowns. According to the principle of the least-square method, the following extreme value condition should be satisfied

$$\frac{\partial E(x,y)}{\partial a(x,y)} = 0, \ \frac{\partial E(x,y)}{\partial b(x,y)} = 0, \ \frac{\partial E(x,y)}{\partial c(x,y)} = 0 \ \frac{\partial E(x,y)}{\partial d(x,y)} = 0, \ \frac{\partial E(x,y)}{\partial e(x,y)} = 0 \tag{4}$$

In this way, five unknowns can be resolved simultaneously from these equations. Solve Eq. (4) to get the following matrices

$$X = U^{-1}Q \tag{5}$$

Where

$$U = \begin{bmatrix} N & \sum_{n=1}^{N} c_{1,n} & \sum_{n=1}^{N} s_{1,n} & \sum_{n=1}^{N} c_{2,n} & \sum_{n=1}^{N} s_{2,n} \\ \sum_{n=1}^{N} c_{1,n} & \sum_{n=1}^{N} c_{1,n}^2 & \sum_{n=1}^{N} c_{1,n}s_{1,n} & \sum_{n=1}^{N} c_{1,n}c_{2,n} & \sum_{n=1}^{N} c_{1,n}s_{2,n} \\ \sum_{n=1}^{N} s_{1,n} & \sum_{n=1}^{N} c_{1,n}s_{1,n} & \sum_{n=1}^{N} s_{1,n}^2 & \sum_{n=1}^{N} c_{2,n}s_{1,n} & \sum_{n=1}^{N} s_{1,n}s_{2,n} \\ \sum_{n=1}^{N} c_{2,n} & \sum_{n=1}^{N} c_{1,n}c_{2,n} & \sum_{n=1}^{N} c_{2,n}s_{1,n} & \sum_{n=1}^{N} c_{2,n}^2 & \sum_{n=1}^{N} c_{2,n}s_{2,n} \\ \sum_{n=1}^{N} s_{2,n} & \sum_{n=1}^{N} c_{1,n}s_{2,n} & \sum_{n=1}^{N} s_{1,n}s_{2,n} & \sum_{n=1}^{N} c_{2,n}s_{2,n} & \sum_{n=1}^{N} s_{2,n}^2 \end{bmatrix}$$

$$X = [a(x,y) \ b(x,y) \ c(x,y) \ d(x,y) \ e(x,y)]^T$$

$$Q = [\sum_{n=1}^{N} I_n \ \sum_{n=1}^{N} I_n c_{1,n} \ \sum_{n=1}^{N} I_n s_{1,n} \ \sum_{n=1}^{N} I_n c_{2,n} \ \sum_{n=1}^{N} I_n s_{2,n}]^T \tag{6}$$

Where $c_{i,n} = \cos\delta_{\lambda i,n}$ and $s_{i,n} = \sin\delta_{\lambda i,n}$ (i = 1, 2).

We prefer to the five-step phase-shifting algorithm. By taking $N=5$ and solving Eq. (5) through five interferograms with arbitrary phase shift, the wrapped phase of $\varphi_{\lambda 1}$ and $\varphi_{\lambda 2}$ can be obtained by

$$\varphi_{\lambda 1} = \arctan(-\frac{c(x,y)}{b(x,y)}), \quad \varphi_{\lambda 2} = \arctan(-\frac{e(x,y)}{d(x,y)}) \tag{7}$$

Thus, the phase of synthetic beat wavelength can be calculated by

$$\varphi = \varphi_{\lambda 1} - \varphi_{\lambda 2} \tag{8}$$

Finally, the thickness of the phase object can be obtained by Eq. (11) in Ref. 1.

To demonstrate the feasibility of the proposed method, the simulated experiments are carried out. Five-frame interferograms with size of 257 × 257 pixels are generated, in which the background is $a=120\exp(-0.05(x^2 + y^2))$ and the modulation amplitudes are $b_{\lambda 1}=50\exp(-0.04(x^2 + y^2))$ and $b_{\lambda 2}=60\exp(-0.05(x^2 + y^2))$, at 532 and 632.8 nm, respectively. The synthetic beat wavelength is equal to 3.3398 $\mu m$. A spherical cap with the height of $h_{set} = 480 \times 10^{-6}(x^2 + y^2)$, in which $-1.28\,mm \leq x, y \leq 1.28\,mm$ is employed as the measured object. In this way, even though the retrieved phases $\varphi_{\lambda 1}$ and $\varphi_{\lambda 2}$ are wrapped, due to that the synthetic wavelength is much larger, the phase of synthetic beat wavelength is not wrapped, leaving out the process of phase unwrapping. As the phase shift can be chosen randomly, Fig. 1 respectively shows five interferograms of the object with phase shifts of $\delta_{\lambda 1,1}=0$, $\delta_{\lambda 1,2}=\pi/3$, $\delta_{\lambda 1,3}=4\pi/5$, $\delta_{\lambda 1,4}=7\pi/6$, $\delta_{\lambda 1,5}=8\pi/5$ at $\delta_{\lambda 1,n}$, and the phase shifts of $\delta_{\lambda 2,n}$ are correspondingly set as $\delta_{\lambda 1,n}\lambda_1/\lambda_2$. Then the wrapped phases of each single wavelength can be calculated from these phase-shifting interferograms by using Eq. (7), as shown in Figs. 2(a) and 2(b). Figs. 3(a)-(b) show the three-dimensional(3D) display of wrapped phase of Figs. 2(a) and 2(b), respectively. Thus, the continuous phase of synthetic beat wavelength can be obtained by using Eq. (8), as depicted in Fig. 3(c). Then the height of the measured object can be determined by Eq. (11) in Ref. 1, of which the mean square error between the theoretical height and the calculated one with the proposed approach is 0.67 nm.

The proposed method combines the generalized phase-shifting algorithm and the least-square algorithm to extract the phase distributions with high accuracy from randomly phase-shifted dual-wavelength interferograms, which has characteristics of less time-consuming and simple operation.

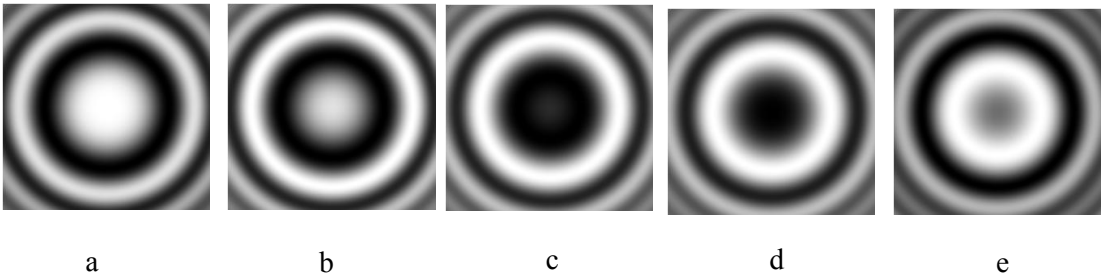

a    b    c    d    e

Fig. 1 Simulated interferograms with different phase shifts: (a) 0 at 532 and 632.8 nm; (b) π/3 at 532 nm; (c) 4π/5 at 532 nm; (d) 7π/6 at 532 nm; (e) 8π/5 at 532 nm.

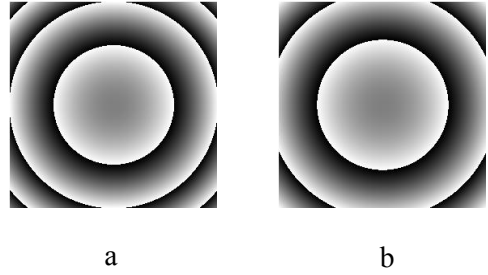

a b

Fig. 2 Wrapped phases: (a) at 532 nm and (b) at 632.8 nm.

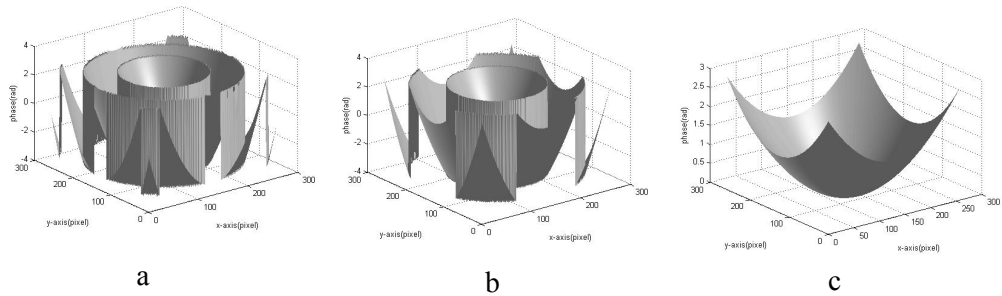

a b c

Fig. 3 3D display of the phases: (a) at 532 nm; (b) at 632.8 nm; (c) at synthetic beat wavelength

Canlin Zhou's email address is canlinzhou@sdu.edu.cn